# Boosting the clean energy transition through data science

Fronzetti Colladon, A., Pisello, A. L., & Cabeza, L. F.

# Boosting the clean energy transition through data science

*Andrea Fronzetti Colladon[a], Anna Laura Pisello[a], Luisa F. Cabeza[b]*

[a] Department of Engineering, University of Perugia, Via G. Duranti 93 – 06125 Perugia, Italy

[b] GREiA Research Group, Universitat de Lleida, Pere de Cabrera 3, Lleida, Spain

The demand for research supporting the development of new policy frameworks for energy saving and conservation has never been more critical. As climate change accelerates and its impacts become increasingly severe, the need for sustainable and resilient socioeconomic systems is increasingly pressing (Aldieri et al., 2021; Cabeza & Ürge-Vorsatz, 2020). The strategic worldwide importance of energy transition and environmental sustainability cannot be understood without the profound knowledge of the context of human well-being and societal and climate resilience. As growing environmental challenges occur, the integration of big data becomes increasingly crucial for promptly detecting predictable phenomena, and finally driving these transitions. Advanced analytics and collective intelligence can now disclose the power of big data to reach up specific insights in drawing dynamic energy consumption patterns, towards the effective implementation of dedicated energy policies, and the exploitation of new sustainable technologies. The timely identification of the driving trends may be carried out by leveraging vast datasets, useful to predict future consumption, and tailor solutions to enhance efficiency and sustainability, while minimising environmental risks and societal vulnerability. For instance, real-time data analysis can optimize the operation of renewable energy sources, leading to a more stable and reliable energy grid.

Additionally, understanding consumer behavior through data analytics allows for more targeted and effective energy-saving initiatives (Siano, 2014) while keeping the same wellbeing and environmental comfort targets.

The significance of big data extends to evaluating and improving energy policies. Policymakers can indeed take advantage of data-driven insights to conceive, design, and implement effective policies that promote sustainable behaviors to rationalize energy use and reduce carbon footprints. At the same time, data analytics may quantitatively disclose the effectiveness of a variety of subsidies, fiscal incentives, and regulatory frameworks and specific measures, supporting governmental actions towards a maximum exploitation. This virtuous feedback loop is in fact essential for driving progressive improvement and human adaptation in the face of continuously evolving environmental and societal needs.

In response to this global challenge, this special issue seeks to explore how advances in Artificial Intelligence (AI) and Data Science can drive the energy transition and enhance environmental sustainability.

The digital economy has significantly impacted the energy transition, enabling more efficient energy use and fostering the adoption of green technologies (W. Li et al., 2023; Shahbaz et al., 2022). Recent technological advancements have revolutionized our ability to make data-driven decisions.

High-quality online data sources offer a unique opportunity to understand how public awareness and citizens' behavior are shaped by policies and media coverage of sustainable energy solutions (Bradshaw et al., 2022; Brown & Sovacool, 2017; Romanova, 2021; Winfield & Dolter, 2014). This understanding is crucial for developing strategies that encourage the adoption of green technologies and improve energy conservation.

As decision-makers increasingly rely on information related to climate change, global warming, and green technology advancements, processing and interpreting large volumes of data becomes indispensable (Fabiani et al., 2023; Piselli et al., 2022; Thomas et al., 2022). Moreover, big data analysis can provide insights into how companies interact with each other and with consumers. These insights can inform the development of more effective business strategies and support the transition to sustainable energy practices (Barchiesi & Fronzetti Colladon, 2021; Vestrelli et al., 2024).

This special issue includes 10 articles addressing the mentioned issues from heterogeneous perspectives.

Understanding public perception and the role of media in shaping attitudes toward energy transition policies is crucial for effective policy implementation. Kim et al. (2024) explore the attitudes of the media and stakeholders towards energy transition policies in Korea. By analyzing online news articles and user comments using advanced machine learning techniques, the study reveals that intense negative emotions dominate public discourse, often fueled by politically charged individuals. The findings suggest that news framing can significantly mitigate emotional intensity and enhance public deliberation.

The work of Vestrelli et al. (2024) focuses on firm communication and investigates the impact of climate risk disclosure on firm market value. This study finds a positive relationship between climate risk disclosure and firm value using text mining and social network analysis on company reports and transcripts of earning conference calls. However, this positive relationship could turn negative when companies face tougher scrutiny due to growing concerns regarding their environmental impact.

Four papers in this issue showcase the potential of machine learning techniques in addressing diverse energy-related challenges. These studies demonstrate the efficacy of machine learning in tasks ranging from market analysis to poverty prediction and policy evaluation within the energy domain. Kocaarslan and Mushtaq (2024) examine the dynamic conditional correlations between municipal green bonds and risky assets during the COVID-19 pandemic. Al Kez et al. (2024) leverage machine learning to predict energy poverty by combining satellite remote sensing with socioeconomic data. This approach successfully identifies districts with significant energy poverty, providing a valuable tool for policymakers. The study of Croce et al. (2024) focuses on identifying cleantech firms and analyzing the impact of national policies on the cleantech landscape in Europe. Lastly, Ates et al. (2024) present a novel method for mapping actors involved in energy and mobility system transitions using Named Entity Recognition (NER). This study highlights the increasing complexity of multi-system governance processes and the pivotal role of energy system actors in driving these transitions.

Two papers discuss how digital technologies and the digital economy impact the clean energy transition. In particular, Luo et al. (2024) investigate countries' path-dependent nature of renewable energy product diversification by using the product space theory. They find that the digital economy plays a crucial role in breaking these path dependencies. Wang et al. (2024) examine the impact of AI on high-quality energy development (HED) across 30 provinces in China. The authors find that AI significantly boosts HED, especially in regions with a well-developed digital economy.

The last two papers included in this special issue provide complementary insights into the complexities of addressing climate challenges within the energy landscape. Ahamed et al. (2024) consider the global oil and gas extraction network and construct a comprehensive dataset to analyze industry cooperation and competition. They reveal significant joint ownership among companies indicative of strategic alignment to obstruct climate policies. Li et al. (2024) employ Bayesian Networks to analyze interdependencies among technologies within decarbonization pathways, highlighting the importance of early transitions in the building sector, consistent deployment of district heating, and the critical role of bioenergy with carbon capture and storage in mitigating emissions.

The entire special issue demonstrates, through quali-quantitative approaches supported by advanced analytics, the pivotal role of cross-disciplinary methods and tools such as AI, text mining, social network analysis, and machine learning in accelerating the clean energy transition.

**Acknowledgments**

We thank the editors of "Energy Policy" for allowing us to realize this special issue, particularly Prof. Leila Dagher, for her continued and invaluable support throughout the editorial process. We also wish to express our sincere appreciation to the numerous reviewers whose dedication, constructive criticism, and insightful suggestions have significantly enhanced the quality of all submitted papers.

**Declaration of generative AI in scientific writing**

While preparing this work, the authors used ChatGPT and Grammarly to improve the language. After using these tools, the authors reviewed and edited the content as needed and take full responsibility for the content of the publication.

analysis of online news. *Renewable and Sustainable Energy Reviews*, *167*, 112792. https://doi.org/10.1016/j.rser.2022.112792

Romanova, T. (2021). Russia's political discourse on the EU's energy transition (2014–2019) and its effect on EU-Russia energy relations. *Energy Policy*, *154*, 112309. https://doi.org/10.1016/j.enpol.2021.112309

Shahbaz, M., Wang, J., Dong, K., & Zhao, J. (2022). The impact of digital economy on energy transition across the globe: The mediating role of government governance. *Renewable and Sustainable Energy Reviews*, *166*, 112620. https://doi.org/10.1016/j.rser.2022.112620

Siano, P. (2014). Demand response and smart grids—A survey. *Renewable and Sustainable Energy Reviews*, *30*, 461–478. https://doi.org/10.1016/j.rser.2013.10.022

Thomas, M., DeCillia, B., Santos, J. B., & Thorlakson, L. (2022). Great expectations: Public opinion about energy transition. *Energy Policy*, *162*, 112777. https://doi.org/10.1016/j.enpol.2022.112777

Vestrelli, R., Fronzetti Colladon, A., & Pisello, A. L. (2024). When attention to climate change matters: The impact of climate risk disclosure on firm market value. *Energy Policy*, *185*, 113938. https://doi.org/10.1016/j.enpol.2023.113938

Wang, B., Wang, J., Dong, K., & Nepal, R. (2024). How does artificial intelligence affect high-quality energy development? Achieving a clean energy transition society. *Energy Policy*, *186*, 114010. https://doi.org/10.1016/j.enpol.2024.114010

Winfield, M., & Dolter, B. (2014). Energy, economic and environmental discourses and their policy impact: The case of Ontario′s Green Energy and Green Economy Act. *Energy Policy*, *68*, 423–435. https://doi.org/10.1016/j.enpol.2014.01.039

9